# How causal inference concepts can guide research into the effects of climate on infectious diseases

Laura Andrea Barrero Guevara[1,2], Sarah C. Kramer[1], Tobias Kurth[2], Matthieu Domenech de Cellès[1]

1. Max Planck Institute for Infection Biology, Infectious Disease Epidemiology group, Charitéplatz 1, Campus Charité Mitte, 10117 Berlin, Germany
2. Institute of Public Health, Charité – Universitätsmedizin Berlin, Charitéplatz 1, 10117 Berlin, Germany.

Corresponding author: Dr. Matthieu Domenech de Cellès, domenech@mpiib-berlin.mpg.de

Keywords: causal inference; climate; weather; infectious diseases; seasonality

**Abstract**

A pressing question resulting from global warming is how infectious diseases will be affected by climate change. Answering this question requires research into the effects of weather on the population dynamics of transmission and infection; elucidating these effects, however, has proven difficult due to the challenges of assessing causality from the predominantly observational data available in epidemiological research. Here, we show how concepts from causal inference—the sub-field of statistics aiming at inferring causality from data—can guide that research. Through a series of case studies, we illustrate how such concepts can help assess study design and strategically choose a study's location, evaluate and reduce the risk of bias, and interpret the multifaceted effects of meteorological variables on transmission. More broadly, we argue that interdisciplinary approaches based on explicit causal frameworks are crucial for reliably estimating the effect of weather and accurately predicting the consequences of climate change.

**Introduction**

A key question ensuing from global warming is how climate change may impact the population dynamics of infectious diseases (1–3). Indeed, observations of large climatic variability in the distribution and seasonality of multiple infectious diseases worldwide—including significant causes of death like malaria (4), cholera (5), and influenza (6)—suggest that many pathogens are sensitive to environmental conditions such that climate change could modify their ecology and epidemiology. More broadly, the fact that phenology—the seasonal timing of events—is a near-universal feature of life on Earth (7) shows that climate is a central determinant of biological processes. Predictive studies, based on numerical simulations combining models of global climate and infectious diseases under different scenarios of greenhouse gas emissions, suggest that climate change affects many infections. These include infections with indirect transmission through intermediate, climate-sensitive stages involving a vector or the environment. Examples of the former include mosquito-borne diseases like malaria and dengue and the latter water-borne diseases like cholera, all predicted to shift their geographical range under continued global warming (8–10). Although fewer studies have focused on directly transmitted pathogens, climate change could also alter the transmission dynamics of respiratory syncytial viruses in the US and Mexico (11) and varicella-zoster viruses in Mexico (12). Although such predictions cannot yet be evaluated, earlier research documented the impact of past climate warming (3), for example, the increased altitudinal range of malaria in the highlands of Ethiopia and Colombia (13) and the increased risk of *Vibrio* disease in northern Europe, coinciding with the warming of the Baltic Sea's surface (14).

A prerequisite to predicting the long-term consequences of climate change is to elucidate the effect of weather on infectious diseases. Even though this effect (and the resulting "calendar of epidemics" (15)) has long been observed, it remains incompletely characterized for many infectious diseases, with persisting uncertainties about the direct causes and mechanisms for even well-researched pathogens like influenza viruses (16,17). Perhaps the most robust evidence for this effect is afforded by experimental studies, which demonstrate that environmental variables like temperature and humidity tightly modulate transmission parameters (such as pathogen survival time or infectivity) of many infections. These include viral infections such as influenza (18) and COVID-19 (19), bacterial infections such as cholera (20) and invasive pneumococcal diseases *(21)*, and parasitic infections such as malaria (22,23). Although such evidence is useful for population-based research (in particular for postulating causal environmental factors), it remains too limited for estimating the causal impact of weather at the scale of human populations for at least three reasons. First, the endpoints measured in studies that experiment only on pathogens—such as survival time—can be challenging to translate into meaningful epidemiological quantities—such as transmissibility. Second, because of differences in infection biology between species, the results from animal studies may not generalize to humans (24). Third, experimental studies cannot recapitulate all the mechanisms whereby weather affects infection, especially those operating at the population level—for example, weather causing behavioral changes resulting in seasonal changes in social contacts (25). Hence, observational studies remain necessary to estimate the multifaceted effects of weather on human infectious diseases.

A well-known shortcoming of observational studies is their proneness to misidentify causes because found associations do not always imply causation for observational data. This problem is also expected when inferring the effect of weather, which is characterized by meteorological variables generally highly correlated with one another and potentially many other seasonal causes of infectious diseases. Causal inference—the sub-field of statistics aiming at inferring causes from observational data—offers a principled approach to tackle these issues and strengthen evidence in observational research (26,27). Causal inference frameworks and their tools are increasingly used to analyze data and guide study design in epidemiology (28) and beyond (29,30). The impact of such tools is illustrated by the fact that causal frameworks like target emulation trials may provide evidence as robust as that from randomized trials (31), thereby expanding the scope of observational research for answering causal questions.

Despite these advances, the use of causal methods—in the form of mechanistic models or statistical models based on causal reasoning—remains limited in the field of weather- or climate-infectious disease research (32). To substantiate this claim, we built on an earlier review (33) and re-analyzed 33 studies that used time-series regression models to assess the association between weather and dengue, influenza, cholera, or malaria cases (Table S1). Although more causal methods for time series analysis exist (34,35), the standard time-series design remains widely used, as evidenced by numerous recent applications for SARS-CoV-2 (36). Four of the 33 studies had an explicitly predictive objective and did not address causality (37–40). Of the remaining 29 studies that addressed causality, only one derived the statistical model from theoretical causal reasoning (*i.e.*, based on a mechanistic model) (41). By contrast, the other 28 studies neither explicitly mentioned causal reasoning to formulate their research question nor used causal graphs for study design or statistical analysis. This lack of causal methods thus suggests the scope for strengthening evidence in this field.

In this perspective, we aim to show how the application of a causal inference framework can help research into the effects of weather—and, by extension, climate—on infectious diseases. Emphatically, our focus is conceptual, and we refer to other excellent reviews for methodological aspects (42,43). Throughout the perspective, we illustrate the different causal inference concepts through a series of vignettes, all based on a simple infectious disease model described below.

**Illustrative infectious disease model**

To illustrate the different causal inference concepts, we formulated a simple mathematical model representing the population-level dynamics of an acute infection spread by direct contact between susceptible and infected hosts. Infection-derived immunity was assumed to be transient, resulting in a risk of re-infection in previously infected hosts. Although we do not focus on a specific pathogen in the rest of this perspective, this model—known as SIRS in the field of infectious disease modeling (44)—may be considered realistic for respiratory viruses with short generation times like influenza viruses (45), respiratory syncytial viruses (46), and SARS-CoV-2 (47).

To describe the effect of weather on infection dynamics, we incorporated an environmental model representing the joint causal effect—dictated by physical laws (48,49)—of ambient air temperature and dew point temperature (a measure of absolute humidity) on relative humidity (see glossary in table 1). Specifically, this model captured the opposing effects of dew point temperature and ambient temperature, with relative humidity increasing with increasing dew point temperature but decreasing with increasing temperature. The latter effect is explained by the physical property that, with increasing temperature, the maximum possible moisture content of the air increases exponentially, such that the denominator grows faster than the numerator in the ratio that defines relative humidity.

To link the environmental and transmission models, we assumed a direct, negative effect of temperature and relative humidity on transmission—that is, transmission decreased as either climatic variable increased. Although the underlying biological mechanisms are yet to be fully elucidated, higher temperature is expected to increase the viral inactivation rate (19), and relative humidity to control the physical and chemical properties of virus-laden respiratory droplets. Specifically, lower humidity can increase evaporation and reduce the size of droplets, potentially facilitating the formation of aerosols with long-range transmission (17,50). Of note, this pair of meteorological variables has been proposed to provide a consistent and mechanistic explanation for the seasonal epidemiology of influenza viruses (17).

To complete the model specification, we formulated a model representing the causal link between the true and the observed incidence rates—that is, an observation model (Fig. 1). In practice, these two rates may differ because of imperfect sensitivity (*e.g.*, due to asymptomatic infections) or specificity (*e.g.*, due to diagnostic tests cross-reacting with other pathogens) in case detection. Here, we considered the relatively typical scenario of a surveillance system with perfectly specific but incompletely sensitive case detection, resulting in an underestimation of the true incidence. Strategic simplifying assumptions (in particular, of a discrete-time model with fixed generation time and time step of one week; see Text S2) allowed us to represent the full model with a simple causal diagram (Fig. 1), also called directed acyclic graph (DAG) in the field of causal inference. We note that a more common causal representation in the field of infectious disease modeling is a flow diagram, displaying all possible epidemiological transitions between model variables with no explicit mention of time. This difference reflects the fact that dynamical modeling studies are not typically framed in a causal inference perspective, even though they also aim to dissect causal mechanisms underlying observations (51).

In the rest of this perspective, we use the full model depicted in Fig. 1 to illustrate four causal inference concepts: Descendants and measurement bias (vignette 1), confounders and confounding bias (vignette 2), natural experiments (vignette 3), and mediators (vignette 4).

**Vignette 1: Descendants, measurement bias, and the intricate association between environmental variables and incidence rate**

Time series regression analysis of observed incidence rates is a frequent study design in environmental epidemiology (33,42). The implicit assumption of such studies is that statistical quantities derived from regression models—typically, regression coefficients—will accurately capture the causal effect of meteorological variables. However, as shown in our causal graph (Fig. 1) and as expected in practice, the causal chain relating weather to observed incidence rates is indirect and complex. A key issue arises from the fact that, although weather directly affects the transmission rate, the observed incidence rate is located two causal links downstream from the transmission rate—with the latter described, in causal inference language, as a descendant (*i.e.*, a consequence) of the former. Of these two causal links, the one relating the transmission and incidence rates may be challenging to recapitulate with regression models because unobserved variables like the size of the population susceptible to infection (inversely related to the population, or herd, immunity, which controls epidemic thresholds) induce non-linearities that may result in marked dissimilarities between these two rates (43,44). Hence, measurement bias—*i.e.*, the bias arising from non-random differences between the targeted and the observed endpoints (52)—may distort causal inference from time series regression models. This potential bias has been recognized in environmental epidemiology, as reflected in recommendations to include additional covariates for capturing temporal variations in population immunity or other long-term trends (42). However, because of the above complexities, such additions, depending on the underlying causal structure and available information, are not guaranteed to reduce measurement bias.

To illustrate, we generated model simulations for a pathogen with low, medium, or high transmissibility (basic reproduction number of 1.25, 2.5, or 5, respectively), with meteorological data from a temperate climate (Lübeck, Germany) resulting in a seasonally-forced transmission rate with a single peak every winter (Fig. 2A). Under the medium-transmissibility scenario (Fig. 2B, middle panel), the epidemiological dynamic displayed annual periodicity, with winter seasonality in the incidence rate that broadly matched that of the transmission rate (Spearman's correlation coefficient: $r_s = 0.63$). In marked contrast, lower transmissibility resulted in biennial epidemics showing little correlation with the seasonal transmission rate ($r_s = 0.07$, Fig. 2B, top panel). This phenomenon—called sub-harmonic resonance (44)—resulted from the higher susceptibility threshold needed to trigger epidemics and the longer time required to replenish the pool of susceptible individuals (via births and waning immunity) to exceed that threshold. Finally, the opposite phenomenon of super-harmonic resonance was observed in the high-transmissibility scenario, which resulted in biannual epidemics (Fig. 2B, bottom panel). These simple numerical experiments illustrate the complex dynamic of infectious diseases and the potent but sometimes counter-intuitive footprint weather—or, for that matter, any other source of seasonal forcing—can have on this dynamic (43).

For each scenario, we next generated 100 replicate time series of observed incidence rates to assess the reliability of time-series regression models. For every replicate, as a control, we first fitted a Negative Binomial generalized additive model with 1–week–lagged weather variables and the susceptible and infected population sizes as covariates and the observed incidence rate as endpoint—the true candidate model for our application, as we show in Text S3. As expected, the estimated causal effects of temperature and relative humidity on the transmission rate were, on average, unbiased for this model (Fig. S1). In practical applications, however, the susceptible and infected population sizes would be unobserved. Therefore, we next fitted a comparable model with a flexible smooth of time to try to capture variations in these unobserved variables. As shown in Fig. 2C, because of measurement bias, estimation performance was, overall, poor. For low transmissibility, the causal effect of temperature on the transmission rate was estimated with a significant bias (mean absolute bias [AB]: 0.08, 40% relative error in comparison to the actual value of –0.2) and imprecision (mean standard error [SE]: 0.07). The bias was even more substantial in the medium- (mean AB: 0.16) and high-transmissibility scenarios (mean AB: 0.22). Because relative humidity had a high correlation with temperature but lower variability, its estimated effect was marred with large uncertainty, which exceeded, on average, the absolute effect size in every scenario (mean SE: 0.21–0.35, to be compared with the true effect size of –0.2, Fig. S1).

Although not intended to be exhaustive, this simple simulation study echoes earlier discussions of measurement bias as a significant concern for study designs based on time-series regression, particularly when population immunity—portrayed as the "dark matter" of epidemic dynamics (53)—varies over fast time scales (33,42). Relating to the central thread of this perspective, we note that causal reasoning—particularly the causal diagram representing the effect of weather—allowed us to identify, *a priori*, the relevant theoretical issues and propose simple numerical experiments to assess their practical relevance. This vignette thus illustrates the value of causal reasoning not only as a methodological tool but also as a theoretical tool to guide study design and analysis.

**Vignette 2: Confounding bias and how climate variability can masquerade as spatial spread**

Spatial heterogeneities are commonly observed for infectious diseases (54–58). Such heterogeneities can result from two broad classes of mechanisms, depending on whether they involve spatial interactions through the movement of individuals (*i.e.*, spatial spread) or spatial variation in some other variable (*e.g.*, climate) (59). For example, weather may vary along a latitudinal gradient and affect pathogen transmission in multiple locations, resulting in spatial covariance between these locations even in the absence of spatial spread. This common cause of spatial variability may thus result in spurious associations that confound the estimated effect of spatial spread on observed incidence—that is, confounding bias (Fig. S2, Table 1).

To illustrate, we considered a scenario with seasonal transmission forcing by weather but no spatial spread in two distinct countries: one with a definite latitudinal gradient in climate (Colombia (60)) and another with little spatial variability in climate (Spain). We simulated the resulting dynamics of our transmission model and assessed epidemic synchrony (61) across various locations in these two countries. In Colombia, the simulated incidence displayed diverse seasonal patterns that followed a latitudinal gradient broadly matching that of climate. (Fig. 3B). In contrast, the low climatic variability in Spain resulted in tightly synchronous epidemics across the locations (Fig. S3B). Hence, despite the absence of mechanisms causing spatial spread in our model, the shared effect of climate between locations resulted in marked spatial correlation in observed incidence, up to ~250 km in Colombia and more extensively throughout Spain (Fig. 3D and Fig. S3D).

To further characterize this spatial correlation, we estimated the speed of the—spurious—traveling wave under the incorrect assumption of spatial spread being the sole cause of spatial heterogeneity (55). Speed estimates were near infinite in Spain (Fig. S3A, C), suggesting either confounding by climate or extremely strong coupling between the locations (44). In Colombia, however, the speed was estimated at 229 km/month (95% CI [124, 472] km/month) (Fig. 3A and C), a value consistent with that documented for real traveling waves—*e.g.*, 110–320 km/mo for pertussis during 1951–2010 in the US (57) and 150 km/mo for dengue during 1983–1997 in Thailand (62).

In the Supplement (Text S4), we further illustrate the confounding problem by fitting bivariate Normal models to time series of observed incidence in a pair of locations in Colombia (Bogotá and Pasto). In keeping with the results above, we found that a model omitting environmental variables estimated a spurious association between the two locations (Pearson correlation coefficient: 0.68, 95% CI [0.63, 0.71]). By contrast, a model including these variables—and thus controlling for the shared climate between the two locations—correctly revealed the absence of spatial spread (Pearson correlation coefficient: –0.06 [–0.13, 0.03]). As in vignette 1, we also fit a bivariate GAM with smooths of time to try to capture the variations in the susceptible and infected population sizes, which would be unobserved in practice. We found that this model, despite being adjusted for the environmental variables, yielded biased estimates suggesting the presence of spatial spread (Pearson correlation coefficient: 0.61 [0.56, 0.65]). This finding thus reemphasizes the difficulty of analyzing incidence data because of measurement bias (vignette 1). More generally, this vignette demonstrates the need for explicit causal models to disentangle the mechanisms underlying spatial heterogeneities.

**Vignette 3: Climate variability as natural experiments to estimate the individual effect of meteorological variables**

Due predominantly to latitudinal gradients in solar radiation and other factors like altitude and proximity to the sea, the Earth displays a large variability of climates (63). This variability is reflected in the Köppen-Geiger system, which classifies worldwide climates into five main types and 30 sub-types based on seasonal averages of precipitation and temperature (64). Because these different climates exhibit diverse seasonal patterns of variation in weather variables and correlations between them, they may be regarded as a range of "natural experiments," conceptually equivalent to manipulating specific variables to identify their causal effects. More broadly, the strategy of leveraging randomness that occurs naturally in observed data (*i.e.*, quasi-experiments) is increasingly advocated for when inferring causality in predominantly observational research fields like economics (29), ecology (65), and epidemiology (66).

Of particular interest for environmental epidemiological research is the contrast between tropical climates (where temperature generally varies little and thus only slightly affects relative humidity) and temperate climates (where the opposite is typically observed), which may be leveraged to isolate the effects of temperature and relative humidity. To test this hypothesis, we used our illustrative model to run a range of numerical experiments in a pair of locations where this contrast was marked: Lübeck, Germany (53.9 °N latitude, coefficient of variation [standard deviation/mean] of temperature $CV(T_e) = 0.59$, $CV(RH) = 0.07$, $r_s(T_e, RH) = -0.48$) and Bogotá, Colombia (4.7 °N latitude, $CV(T_e) = 0.04$, $CV(RH) = 0.07$, $r_s(T_e, RH) = -0.10$). By back-fitting our transmission model to 100 replicate time series of observed incidence rates it generated, we gauged how well we could estimate the effects of temperature and relative humidity (as well as other model parameters that would be unknown in real-world applications; see Text S5) in the two climates. In Lübeck, as expected for a climate characterized by low RH variability and large RH–Te correlation, the effect of temperature was estimated with more accuracy than that of relative humidity (mean AB of 0.02 and 0.09, respectively, Fig. 4). Of note, these results are reminiscent of those of vignette 1, except that the parameters estimated in this vignette originated from the true causal transmission model and, therefore, did not suffer from measurement bias. In Bogotá, again in keeping with intuition for this climate with low Te variability and almost null RH–Te correlation, the opposite result held, with higher accuracy for relative humidity than for temperature (mean AB of 0.04 and 0.05, respectively).

This simulation study thus suggests the scope for strategic choices regarding a study's location, where the local climate's properties can help estimate the effect of the weather variable of causal interest. Whenever more data are available, an alternative strategy is to use multilevel models, which provide a principled way to pool information while modeling variation across multiple locations (52). Multilevel extensions are now routine for standard regression models but more challenging for the complex—typically non-linear, stochastic, and partially observed (67)—models needed to capture infectious disease dynamics. Nevertheless, recent statistical advances permit the estimation of such multilevel models (68,69), opening an avenue for large-scale dynamical modeling studies that harness information from multiple natural experiments in different climates.

**Vignette 4: Mediation and the direct and indirect causal effects of temperature on transmission**

Weather results from a complex web of causally related environmental variables that can affect pathogen transmission through multiple pathways (34). For instance, as illustrated in our simple causal graph and supported by experimental evidence (19), both temperature and relative humidity may directly affect transmission. However, because temperature also impacts relative humidity, its total causal effect may comprise a direct effect ($T \rightarrow \beta$) and an indirect effect mediated by relative humidity ($T \rightarrow RH \rightarrow \beta$). Importantly, but counter-intuitively, the direct and indirect effects may act in opposite directions depending on the causal relationship between the parent variable (*e.g.*, temperature) and its mediator (*e.g.*, relative humidity) in the environmental model.

To illustrate how these two effects play out in different settings, we simulated our model in Lübeck, Germany, and Pasto, another Colombian city with RH variability higher than in Bogotá. Due to climatic differences causing temperature to be more variable than relative humidity in Lübeck ($CV(T_e) = 0.59$, $CV(RH) = 0.07$) but less variable in Pasto (1.4 °N latitude, $CV(T_e) = 0.07$, $CV(RH) = 0.19$), we hypothesized the total impact of temperature would differ between the two cities. In each city, we considered two scenarios. In the first scenario, we set the two climatic parameters (see legend of Fig. 1) to their baseline values to capture the total effect of temperature—*i.e.*, the direct effect and the indirect effect mediated by relative humidity. In the second scenario, we set the climatic parameter of temperature to 0 to capture only its indirect effect, mediated by relative humidity.

In both cities, because of a combination of two negative effects (of temperature on relative humidity and relative humidity on transmission), higher temperature *increased* transmission through the indirect pathway ($r_s(T_e, \beta) = 0.91$ in Pasto and $0.72$ in Lübeck, Fig. 5A and 5B). In Lübeck, however, the higher temperature variability caused the direct effect to outweigh this indirect effect so that, overall, transmission decreased as temperature increased ($r_s(T_e, \beta) = -0.99$, Fig. 5A). By contrast, the higher variability in relative humidity reversed the total effect in Pasto, where transmission increased with temperature ($r_s(T_e, \beta) = 0.80$, Fig. 5B). Despite this overall positive effect, adjusting for relative humidity revealed the negative direct effect of temperature in Pasto (Partial Spearman's rank correlation coefficient: $r_s(T_e, \beta | RH) = -0.56$, Fig. 5B). Of note, echoing the results of vignette 1, the effects of temperature on the observed incidence rates were less definite because of measurement bias (Fig. S4). Hence, despite identical causal mechanisms, climatic differences resulted in divergent effects of temperature in the two cities.

These conceptual insights have practical implications for interpreting the association between environment and transmission rates. Specifically, when evaluating the effect of temperature in a DAG similar to that in Fig. 1, models adjusting for relative humidity would identify the direct effect of both variables. In contrast, models without adjustment would only identify the total effect of temperature. Hence, the lack of clear causal frameworks may lead to misinterpreting model outputs, a risk earlier described as the "Table 2 fallacy" (70). Hence, this vignette reemphasizes the critical importance of causal reasoning and careful interpretation when probing the effect of climate on infection dynamics.

**Discussion**

In this perspective, we aimed to show how causal inference concepts—such as descendants and mediators, confounding and measurement biases, and quasi-experiments—can guide research into the effects of climate on infectious diseases. Through a series of case studies, we illustrated how such concepts could help assess study design (vignette 1), evaluate the risk of confounding bias (vignette 2), strategically choose a study's location to achieve the set-up of a natural experiment (vignette 3), and interpret the direct and—sometimes paradoxical—indirect effects of meteorological variables on transmission (vignette 4). More broadly, seconding earlier calls in the epidemiological field (28), we argue that explicit causal frameworks are necessary for inferring the effect of weather and subsequently predicting the consequences of climate change on infectious diseases.

Because of this perspective's conceptual focus, we sidestepped the many methodological technicalities that inevitably arise in practice. In particular, the effect of weather on infectious diseases can be more intricate than our simple model suggests. First, more environmental variables may have a direct effect on transmission, such that disentangling their direct and indirect effects may be more challenging than in vignette 4. For example, if one assumes a direct effect of dew point temperature on transmission in the DAG represented in Fig. 1, then a causal mediation analysis would be required to estimate these different effects for all the environmental variables. Second, because of interindividual variability in the period separating exposure from infectiousness (*i.e.*, the latent period), the effect of weather on transmission is expected to be lag-distributed, resulting in a more complex causal diagram than Fig. 1. Third, this effect may be non-continuous and non-monotonic, as illustrated by recent experimental evidence showing a V-shape, threshold association between relative humidity and survival time of coronaviruses (19,71). Fourth, although the causal graph of Fig. 1 is realistic for pathogens causing acute, directly transmitted infections (such as respiratory viruses), the weather may have multiple effects on pathogens with more complex relationships with their human hosts. Such multiple effects are, for example, expected for invasive bacteria with prolonged carrier states (like the pneumococcus (72)), with weather affecting not only transmission of carriage but also progression from carriage to invasive disease (73,74). Finally, the poor correlation between indoor and outdoor meteorological variables (especially observed for temperature and relative humidity (75)) has led to discussions about which measure of weather is more appropriate for causal inference, with indoor data argued to represent the bulk of weather exposures (17,76). This problem may be viewed as another form of measurement bias and treated in a causal framework by modeling the causal link between indoor variables and their outdoor counterparts, with some recent research in this direction (76). More generally, our simple causal framework could be similarly extended to tackle the other complexities listed above and the potential biases illustrated in the vignettes simultaneously.

In conclusion, the growing field of causal inference provides opportunities to strengthen evidence derived from observational data. This perspective thus presents an early effort to cross-fertilize this field with infectious disease epidemiology and climatology, with the ultimate research aims of elucidating how climate affects pathogens and predicting the consequences of climate change. Because phenology is a near-universal feature of life, such research may also lead to new insights into the ecology of infectious diseases.

# Figures

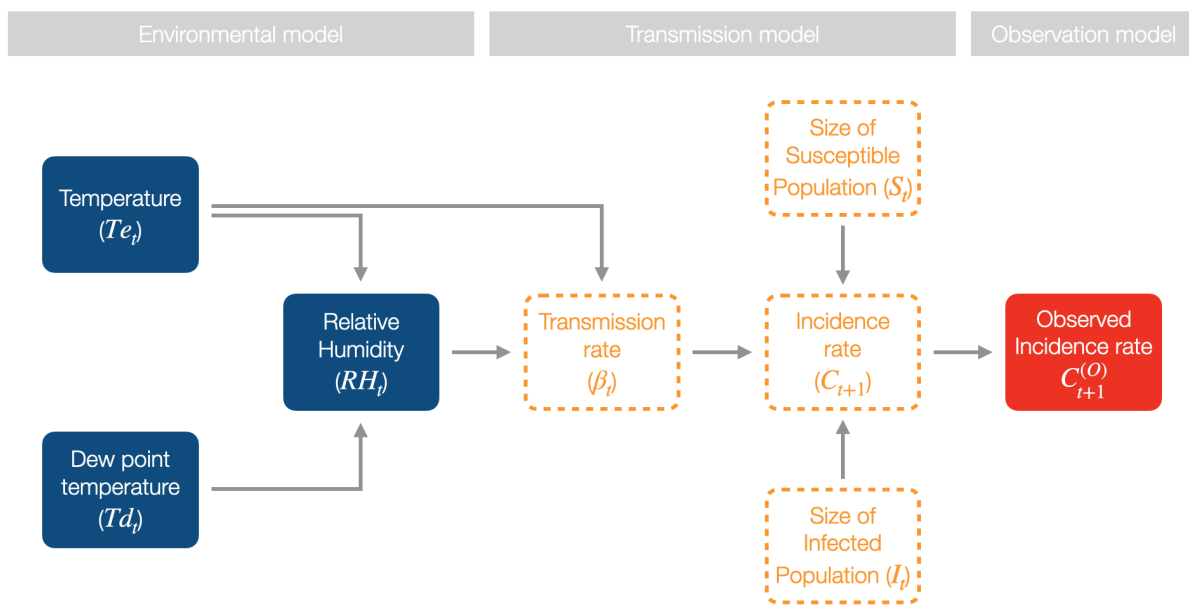

**Figure 1: Causal graph for the illustrative transmission model.** At a given time *t*, physical laws (48,49) dictate that the ambient temperature and the dew point temperature jointly determine the relative humidity of the air. We assume that both temperature and relative humidity directly affect the transmission rate of the pathogen: $\log \beta_t = \log \beta + \delta_{Te}(Te'_t - 1) + \delta_{RH}(RH'_t - 1)$, where $\beta$ represents the average transmission rate, $Te'_t$ and $RH'_t$ the rescaled environmental variables (with unit mean), and $\delta_{Te} = \delta_{RH} = -0.2$ their causal effects on transmission. The causal link between the transmission rate and the incidence rate—defined as the number of new cases per unit of time—is governed by a deterministic non-linear model (here assumed to be a simple SIRS model (44)) representing the pathogen's epidemiological dynamics, in which new infections arise from contact between susceptible and infected individuals. Because of observation error (*e.g.*, imperfect test sensitivity resulting in case under-reporting), the observed incidence rate differs from the actual incidence rate; a stochastic observation model (here assumed to be a Negative Binomial model) represents the causal link between the two rates. Mathematically, this causal diagram translates into a discrete-time model that iterates the epidemic dynamic from one generation of infection to the next (see the supplement for full information about the model's formulation and implementation). Variables surrounded by dashed lines are assumed to be unobserved.

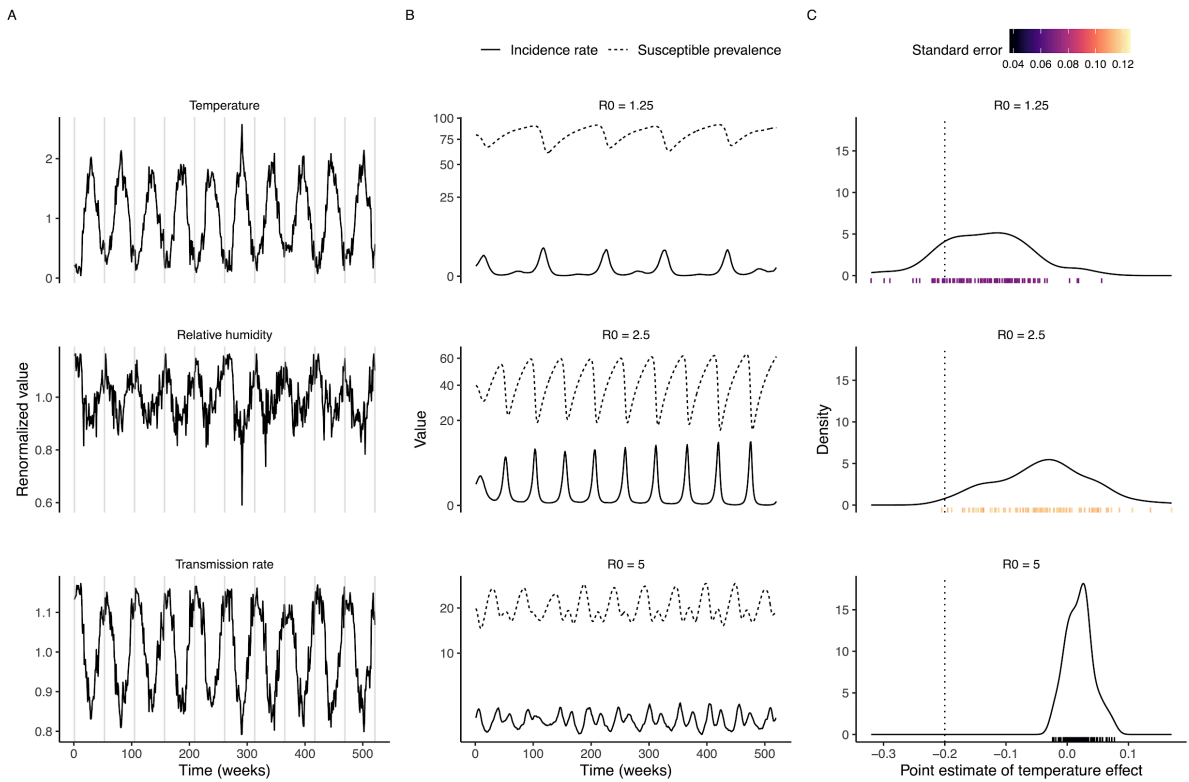

**Figure 2: Measurement bias and the intricate association between environmental variables and incidence rate (vignette 1).** A: Time series of temperature (top panel) and relative humidity (middle panel, both climatic variables renormalized to have unit mean) during 2013–2022 in Lübeck, Germany. The bottom panel displays the resulting seasonal component of the transmission rate, assuming a negative effect of both climatic variables on transmission. In the three panels, the vertical gray lines mark the beginning of every calendar year (week no 1). B: Dynamics of susceptible prevalence ($100 \times S_t/N$, dashed lines) and incidence rate ($100 \times C_t/N$, solid lines) for three different values of the basic reproduction number and an average duration of immunity of 1 yr (other model parameters fixed to the values in Table S3). C: Distribution of point estimates for the effect of temperature estimated from a Negative Binomial GAM regression model fitted to each of 100 replicate time series of the observed incidence rate. The marks on the *x*-axis indicate the point estimates, with the colors representing their standard errors. The vertical dotted line indicates the true effect fixed in the transmission model for generating the observations.

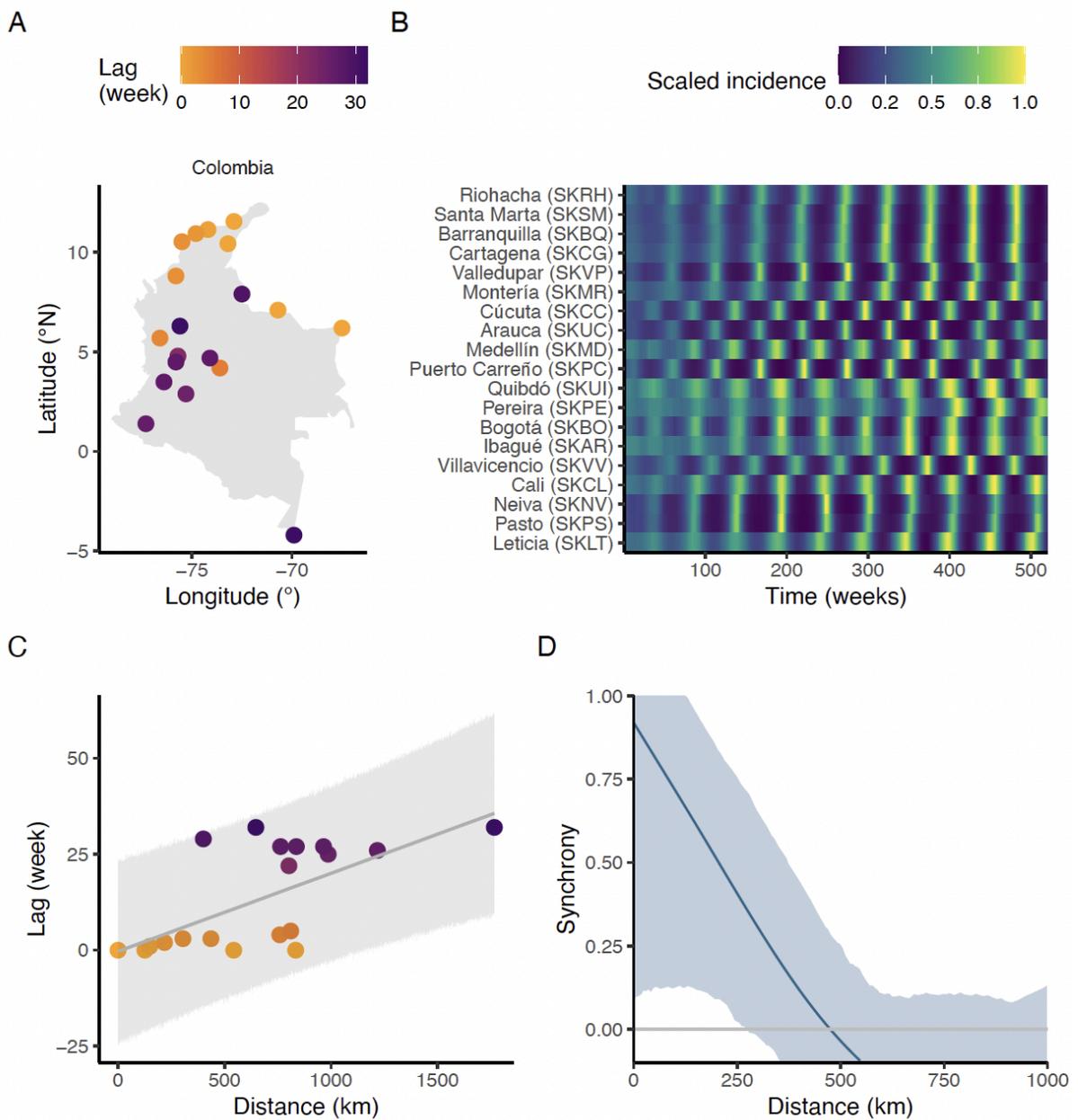

**Figure 3. Confounding bias and how climate variability can masquerade as spatial diffusion in Colombia (vignette 2).** A: Considering 19 locations across Colombia, we B: simulated incidence with no spatial diffusion but a common effect of climate on transmission. C: Relative timing of epidemic peaks (with reference to the northernmost site located near the city of Riohacha) across Colombia. The color indicates the time difference between the epidemic peaks of each site and those of the reference site. The line shows the estimated speed of the traveling wave. D: Pairwise epidemic synchrony between sites. Model parameters: basic reproduction number of 2.5, average duration of immunity of two years, and other parameters as in Table S3.

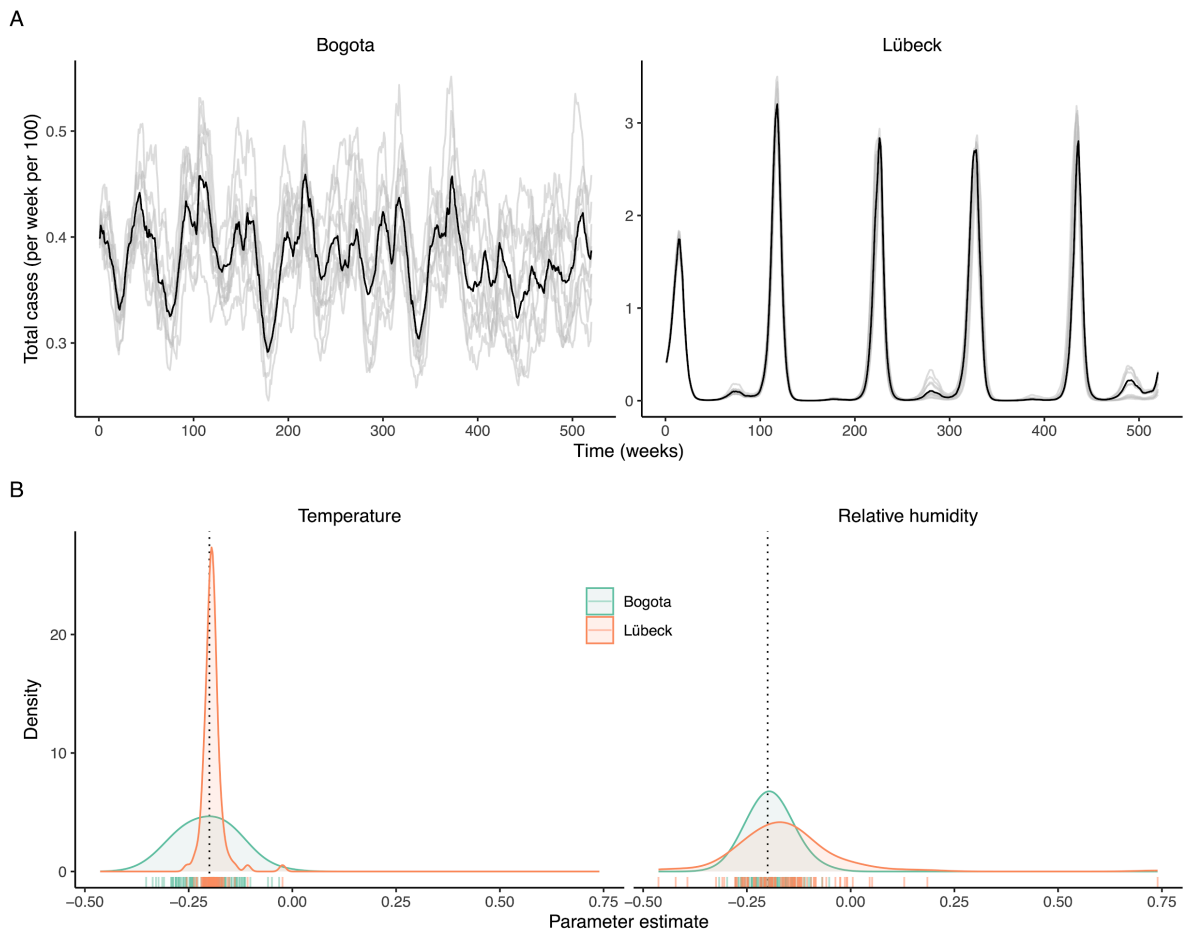

**Figure 4: Climate variability as natural experiments to estimate the individual effect of meteorological variables (vignette 3)**. A: Simulated incidence rate ($C_t$) in Bogotá, Colombia (left panel) and Lübeck, Germany (right panel). Model parameters: basic reproduction number of 1.25, average duration of immunity of one year, and other parameters as in Table S3. In each panel, the black line shows the dynamic from the deterministic transmission model. The grey lines represent simulations (10 displayed out of 100 overall) from a stochastic transmission model, in which noise was added to the transmission rate at every time point (see the Supplement for full details). We implemented this stochastic variant only for this vignette to add realism in the form of model misspecification during estimation. B: Distribution of the 100 maximum likelihood estimates of the effect of temperature (left panel) and relative humidity in Bogotá and Lübeck. These estimates were obtained by direct maximization of the log-likelihood (*i.e.*, trajectory matching) by fitting the misspecified model—in which the transmission model was deterministic and the observation model stochastic—to data generated from the fully stochastic model—in which both the transmission and the observation models were stochastic. The dotted vertical lines indicate the true parameter value (–0.2) fixed in all model simulations. See Text S5 for further details about the estimation procedure.

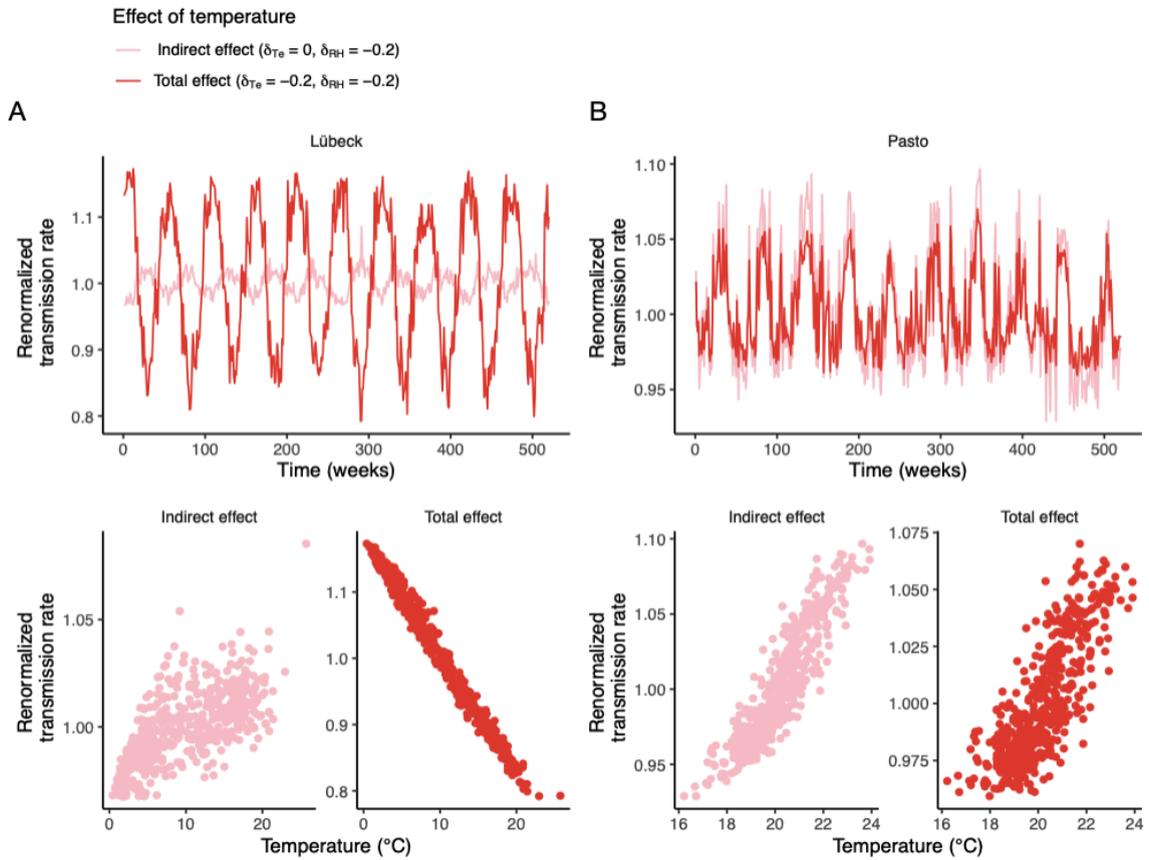

**Figure 5. Mediation and the direct and indirect causal effects of temperature on transmission (vignette 4).** Simulated transmission rate (top panels) in A: Lübeck, Germany, and B: Pasto, Colombia, from models including the total effect of temperature (dark lines) or only the indirect effect of temperature mediated through humidity (light lines). The bottom panels show the indirect effect of temperature through relative humidity on transmission (left panels) and the total effect of temperature (right panels). Model parameters: basic reproduction number of 1.25, average duration of immunity of one year, and other parameters as in Table S3.

# Tables

Table 1. Glossary of environmental, infection dynamics, and causal inference concepts.

| Concept | Definition | Section |
|---|---|---|
| **Environmental concepts** | | |
| Dew Point Temperature ($T_d$) | The temperature at which air must be cooled to become saturated with water vapor and form dew. It is a measurement of absolute humidity. | Illustrative infectious disease model: Environmental model |
| Relative Humidity (RH) | The ratio of the current moisture content in the air to the maximum moisture content of the air at a specific temperature. | |
| **Infection disease modeling concepts** | | |
| Basic Reproduction Number ($R_0$) | The average number of secondary infections generated by a single infected individual in a completely susceptible population. | Illustrative infectious disease model: Transmission model |
| Transmission rate ($\beta_t$) | The rate at which infected individuals transmit to susceptible individuals at time *t*. | |
| Incidence rate ($C_t$) | The cumulative number of new infections in a defined population per unit of time. | |
| Susceptibles ($S_t$) | The size of the population that is not immune and thus susceptible to contracting infection at time *t*. | |
| Infected ($I_t$) | The size of the population that is infected and can infect susceptible individuals at time *t*. | |
| Recovered ($R_t$) | The size of the population that was previously infected but is no longer infectious and remains immune against re-infection at time *t*. | |
| **Causal Inference concepts** | | |
| Directed Acyclic Graph (DAG) | A causal diagram where the nodes are variables, and the edges are direct causal effects between variables. | Vignettes 1–4 |
| Descendant | A variable Y that is directly or indirectly caused by another variable (X → Y). | Vignette 1 |
| Measurement bias | A bias on the estimated effect of exposure X on outcome Y resulting from the process by which the variables are measured (X* or Y*). | Vignette 1 |
| Confounding bias | A bias on the estimated effect of an exposure X on outcome Y when the exposure and the outcome share a common cause C (X ← C → Y). The common cause is often referred to as confounder. | Vignette 2 |
| Quasi-experiments | Leveraging randomness that occurs naturally in observed data to evaluate causal effects. | Vignette 3 |
| Mediator | A middle variable M through which an exposure X affects an outcome Y (X → M → Y). | Vignette 4 |

**Paper information**

*Data availability statement*. All data and R programming codes are available at https://gitfront.io/r/MDdC/F5m14YVq1642/Causality-Seasonality/

*Declaration of interests.* T.K. reports outside the submitted work, having received research grants from the Gemeinsamer Bundesausschuss (G-BA – Federal Joint Committee, Germany) and the Bundesministerium für Gesundheit(BMG – Federal Ministry of Health, Germany). He further has received personal compensation from Eli Lilly and Company, Teva Pharmaceuticals, TotalEnergies S.E., the BMJ, and Frontiers. M.D.d.C. received postdoctoral funding (2017–2019) from Pfizer and consulting fees from G.S.K. All other authors declare no conflicts of interest.

*Funding*. This work was funded by the Max Planck Society through the core funding of MDdC's Max Planck Research Group at the Max Planck Institute for Infection Biology.